\documentstyle[11pt,newpasp,twoside,epsf]{article}
\def\edcomment#1{\iffalse\marginpar{\raggedright\sl#1\/}\else\relax\fi}
\reversemarginpar

\begin{document}
\title{The study of $\delta$ Sct stars in the transition era from ground--based to
space photometry}
\author{Rafael Garrido}
\affil{Instituto de Astrofisica de Andalucia, CSIC Apdo. 3004, 18080 Granada, Spain}
\author{Ennio Poretti}
\affil{INAF-Osservatorio Astronomico di Brera, Via E. Bianchi 46, 23807 Merate, Italy}

\section{Introduction}

The different research teams involved in the study of $\delta$ Sct stars
have slightly changed their strategy in the past years. The
observational effort to secure worldwide coverage of case studies is
continued, but the  specifications become more severe, especially
about target characterization and frequency resolution.

After the successful launch of the Canadian satellite MOST, which will be the
pioneer of asteroseismology from space, 
the future missions are programmed to properly take into account  the need of 
an adequate frequency resolution: COROT will spent 30 and 150 days (additional
and core programs, respectively) on the target, while EDDINGTON will arrive
up to 1 year.
Such a  requirement is a direct consequence of the observational
results on $\delta$ Sct, $\gamma$ Dor, SPB stars,~... obtained
from ground. It should be noted that without these results (see Poretti 2000
for a review about $\delta$ Sct stars) the scientific
background of the space missions would be much less defined and the risks
of incomplete results (owing to inaccurate selection of targets, insufficient
resolution, underestimate of the influence of the rotation) much higher. 

\section{The increasing number of frequencies detected in FG Vir}

L\'opez de Coca et al. (1984) easily detected the main period of FG Vir on the
basis  of few nights. It can be considered the discovery step.
After that, Mantegazza, Poretti \& Bossi (1994) detected seven frequencies on
the basis of a 14--nights run at the European Southern Observatory. It can be
considered the second step, i.e., the limit for a single--site campaign. 
Owing to its equatorial position,
FG Vir was a suitable candidate for multisite campaign, being possible to observe it
from both hemispheres. Indeed, Breger et al. (1999) confirmed the seven frequencies
and increased the total number of detected frequencies to 24, thanks to a campaign
involving 6 observatories during 40 days. Again, this can be considered the limit
for  a multisite campaign which can be organized with a moderate effort. The next step
was to set up  a campaign not only involving more than 6  sites, bust also spanning a longer 
time baseline. Moreover, the recent results obtained by Mantegazza \& Poretti (2002) 
strongly indicated how  spectroscopy can supply  
useful hints about the inclination of the rotational axis and to propose 
some mode identifications.
Therefore, the 2002--03 campaign on FG Vir has been performed by combining
spectroscopy and photometry over a 2--years time baseline.
The aim is to detect close doublets and investigate in a more powerful way the 
eventual amplitude variations. Such a long--term campaign constitutes the
new fronteer of the ground--based observations.

Figure~1 shows how the 
number of detected frequencies increased by refining the observational approach.
>From Fig.~1 it can be extrapolated that the lowering of the threshold amplitude 
by a factor 10$^{-1}$ (which should be easily attainable from space) will allow us to 
detect hundreds of excited modes. 

\begin{figure}
\begin{center}
\mbox{\epsfxsize=0.5\textwidth\epsfysize=0.5\textwidth\epsfbox{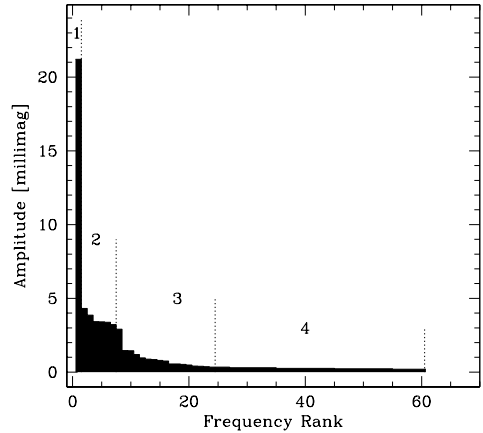}}
\caption{The increasing number of detected frequencies in the FG Vir light curve
on the basis of different observational efforts. 1: L\'opez de Coca et al. (1984);
2:  Mantegazza, Poretti \& Bossi (1994); 3: Breger et al. (1999); 4: Breger 
(2003, priv. comm.)}
\label{fg}
\end{center}
\end{figure}

\section{The preparation of the COROT space mission}

The COROT satellite will monitor targets which must be selected in two fields centered
at  $\alpha=6^h50^m$, $\delta=0$$^{\circ}$  (Anticenter direction),
and $\alpha=18^h50^m$, $\delta=0$$^{\circ}$ (Center direction). The prototype of 
the class, $\delta$ Sct itself, is included in the Center direction field, but it
is considered too evolved to match the goals of the mission, as 
asteroseismic inferences for these stars are greatly complicated by the richness
of excited modes in a narrow frequency range (275 possible modes in a 4~cd$^{-1}$ 
interval; Templeton et al. 1997). Therefore,
we searched for new $\delta$ Sct stars located in the COROT field--of--view which
are unevolved stars (high--priority for COROT) or little evolved ones
(low--priority stars). Efforts have been concentrated 
especially in the galactic Center direction, as there are no many solar--like candidates in
such a direction and target allocation is easier. 

As a result, we actually scanned the solar neighbourhood; $\delta$ Sct stars are
commonly found and the incidence of variability is around 23\% (13 variables out
of 57 candidates) for stars located
in the lower part of the instability strip, i.e., the region  between the
ZAMS and the blue hooks of the evolutionary tracks (Poretti et al. 2003). With such
a percentage, the possibility to find $\delta$ Sct stars in a limited
sky region is reasonable and they can provide a
suitable  basis of asteroseismic targets for any space mission. Figure~2
shows the specific result
obtained for COROT. The fine tuning of the candidates is now  possible and
the ground--based preparatory work can greatly take advantage from the expertise
piled--up in the past decades on the case studies. It is suitable that the
$\delta$ Sct stars which will be 
selected as COROT targets will became case studies for spectroscopy
and/or preliminary photometry. 

\begin{figure}
\begin{center}
\mbox{\epsfxsize=0.7\textwidth\epsfysize=0.7\textwidth\epsfbox{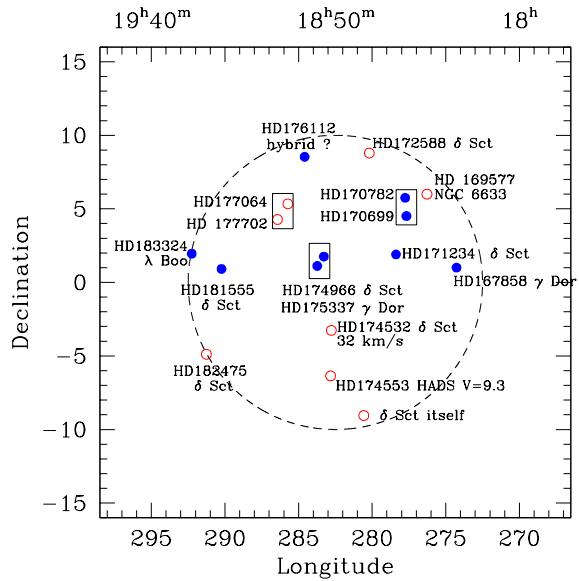}}
\caption{The location of the  stars of interest in the field--of--view
of COROT, Galactic Center direction. Filled circles are unevolved stars, open circles are
evolved ones. Three couples of stars can be observed together in the Seismo CCD,
shown as small rectangular boxes. The type of variability or some physical characteristic
are also reported.}
\end{center}
\end{figure}

\section{Colour information from space missions}
When performing photometry in different colours it is possible
to calculate solutions for each passband. The
phase differences  and the amplitude ratios
will be then used to discriminate the different spherical harmonics of a non-radially 
pulsating star, therefore helping to identify the oscillation mode (Garrido 2000).

 The explanetary camera of the COROT space mission consists on a CCD with a bi-prism 
in front of it giving coloured spots on the detector (see Vauclair in these proceedings 
for details). The three colours basically supplied by the camera are sufficient 
to distinguish the different spherical degrees $\ell$  of the corresponding 
radial or non-radial excited periods, making therefore possible the mode identification of 
several types of pulsating stars, including $\delta$ Scuti stars. Physical information 
derived 
from these coloured data also includes limb-darkening coefficients, non-adiabatic 
phase lags (see Garrido 2003 for details) and, for some specific cases, clues 
concerning stellar rotation (see Daszynska-Daszkiewicz et al. 2002 for details).


EDDINGTON is an ESA space mission devoted to perform asteroseismology and 
planet detection by the method of transits. The present status of the mission 
consists of three telescopes: one will receive white light,  while the other are
equipped with two  
coloured filters blocking only a small (5 to 10 \%) fraction of the light
from the red (Ultrawide Blue Filter) or from the blue (Ultrawide Red Filter). 
Differences between the white light and 
these two coloured filters give narrow enough photometric bands to be useful for 
mode identification, as shown in detail in Deeg \& Garrido (2003). In this way no many 
photons are lost, as required by the original mission specifications. Obviously the 
system is valid only for stars showing relatively high S/N photometric amplitudes 
(see Garrido 2003 for details). $\delta$ Scuti will have sufficient S/N ratios as to 
easily derive phases and amplitudes for the different colours. 



\begin{references}
\reference Breger, M., Handler, G., Garrido, R., et al. 1999, \aap, 349, 225
\reference L\'opez de Coca, P., Garrido, R., Costa, V., \& Rolland, A.
1984, IBVS 2465
\reference Daszynska-Daszkiewicz, J., Dziembowski, W.A. 2002 
\& Goupil, M.J. \aap, 392, 151.
\reference Deeg, H., \& Garrido, R., 2003, Eddington Internal Report.
\reference Garrido, R., 2000, in ``Delta Scuti and Related Stars",
Eds. M. Breger and M.H. Montgomery, ASP Conf. Ser., vol. 210, p.~67
\reference Garrido, R., 2003, 2nd Eddington workshop, Palermo.
\reference Mantegazza, L., \& Poretti, E., 2002, \aap, 396, 911 
\reference Mantegazza, L., Poretti, E., \& Bossi, M. 1994, \aap, 287, 95
\reference Poretti, E., 2000, in ``Delta Scuti and Related Stars",
Eds. M. Breger and M.H. Montgomery, ASP Conf. Ser., vol. 210, p.~45
\reference Poretti, E., Garrido, R., Amado, P.J., et al., 2003, \aap, 406, 203
\reference Templeton, M., McNamara, B.J., \& Guzik, J.A., 2003, \aj, 114, 1592
\reference Zima W., Heiter U., Cottrell P.L., Lehmann H., Mathias P., Poretti E., \& 
Breger M., 2003,
in ``Asteroseismology Across the HR Diagram", Eds. M.J. Thompson, M.S. Cunha, M. Monteiro, Kluwer
Academic Publ., P489
\end{references}
\end{document}